\documentclass[aps,prl,twocolumn,groupedaddress]{revtex4}

\usepackage[mathcal]{eucal}
\usepackage{makeidx}
\usepackage[nice]{nicefrac}
\usepackage[tight]{units}
\usepackage[dvips]{graphicx}

\begin{document}
\preprint{}
% Use the \preprint command to place your local institutional report
% number in the upper righthand corner of the title page in preprint mode.
% Multiple \preprint commands are allowed.
% Use the 'preprintnumbers' class option to override journal defaults
% to display numbers if necessary
%\preprint{}

%Title of paper
\title{Spin correlations in the paramagnetic phase and ring exchange
in La$_2$CuO$_4$}

% repeat the \author .. \affiliation  etc. as needed
% \email, \thanks, \homepage, \altaffiliation all apply to the current
% author. Explanatory text should go in the []'s, actual e-mail
% address or url should go in the {}'s for \email and \homepage.
% Please use the appropriate macro foreach each type of information

% \affiliation command applies to all authors since the last
% \affiliation command. The \affiliation command should follow the
% other information
% \affiliation can be followed by \email, \homepage, \thanks as well.
%option in \documentclass). \noaffiliation is required (may also be
%used with the \author command).
%\collaboration can be followed by \email, \homepage, \thanks as well.

\author{A.M.~Toader{$^1$}, J.P.~Goff{$^1$}, M.~Roger{$^2$}, N.~Shannon{$^2$},
J.R.~Stewart{$^3$} and M.~Enderle{$^3$}}
\date{\today}
\affiliation{ {$^1$} Department of Physics, University of Liverpool,
Oliver Lodge
Laboratory, Liverpool L69 7ZE, United Kingdom.\\
{$^2$} Service de Physique de l'Etat Condens\'e, Commissariat \`a
l'Energie Atomique,\\
Centre d'Etudes de Saclay, 91191 Gif sur Yvette Cedex, France.\\
{$^3$} Institut Laue-Langevin,156X, 38402 Grenoble Cedex, France.}

\begin{abstract}
Spin correlations in the paramagnetic phase of La$_2$CuO$_4$ have
been studied using polarized neutron scattering, with two
important results. First, the temperature dependence of the
characteristic energy scale of the fluctuations and the amplitude
of the neutron structure factor are shown to be in quantitative
agreement with the predictions of the quantum non-linear sigma
model. Secondly, comparison of a high-temperature series expansion
of the equal-time spin correlations with the diffuse neutron
intensity provides definitive experimental evidence for ring
exchange.
\end{abstract}

% insert suggested PACS numbers in braces on next line
% insert suggested keywords - APS authors don't need to do this
%\keywords{}

%\maketitle must follow title, authors, abstract, \pacs, and \keywords
\maketitle

\narrowtext

Heisenberg was the first to realize that strong effective spin
interactions arise from the principle of the indistinguishability
of the particles\cite{Heisenberg}. Dirac generalized this concept,
in the context of Group Theory, to include higher order
interactions\cite{Dirac}. Multi-particle  exchange dominates the
physics of the quantum solid $^3$He\cite{RDH} but, surprisingly,
is generally not taken into account for electronic magnetic
materials. The most powerful technique for exploring exchange
interactions is the study of excitations from the ordered phase
using inelastic neutron scattering. However, ambiguities in the
interpretation of magnon dispersion curves sometimes mean that
higher order terms remain hidden. By employing the independent
approach of studying the instantaneous spin correlations in the
paramagnetic phase, we obtain complementary information that
enables a better understanding of the exchange mechanism. In this
letter we describe studies of the diffuse magnetic scattering from
La$_2$CuO$_4$ which provide compelling, quantitative evidence for
the existence of four-particle cyclic exchange.

La$_2$CuO$_4$ is of great intrinsic interest both as the parent
compound of a canonical high-temperature superconductor, and as a
very good realization of a two-dimensional quantum Heisenberg
antiferromagnet (2DQHAF). Magnetic Raman experiments\cite{Sugai},
infrared absorption studies\cite{Perkins,Lorenzana} and inelastic
neutron scattering measurements\cite{Coldea} show definitively the
inadequacy of the nearest-neighbour Heisenberg model, and suggest
the possibility that four-particle exchange may be significant.

In an important series of experiments to study the diffuse
magnetic scattering from La$_2$CuO$_4$ using unpolarized
neutrons\cite{Birgeneau} the temperature dependence of the
magnetic correlation length was found to agree with the
predictions of the quantum non-linear sigma model
(QNL$\sigma$M)\cite{Chakravarty}. However, the observed amplitude
shows dramatic deviations from the predictions of this
theory\cite{Birgeneau}. The QNL$\sigma$M is the simplest possible
effective action for a 2DQHAF that is compatible with the
long-wavelength spin waves and that does not assume a
spontaneously broken symmetry. Moreover, its predictions should
hold even in the presence of four-spin exchange, as discussed
below. Here we study the dynamical spin correlations above the
N\'{e}el temperature using polarized neutrons, and find complete
agreement of the observed 2D critical fluctuations with the
predictions of the QNL$\sigma$M.

Dirac's approach provides the most transparent theoretical framework
to examine higher-order exchange interactions\cite{Dirac}. His
analysis leads to an effective spin Hamiltonian
\begin{equation}\label{eqn1}
{\cal H}_{eff}=-\sum_\lambda (-1)^{p_\lambda}J_\lambda {\cal
P}_\lambda^\sigma
\end{equation}
where $\lambda$ runs over all possible permutations of spin ${\cal
P}_\lambda^\sigma$ within the symmetric group, $J_\lambda$ is the
exchange energy associated with a given permutation and
$p_\lambda$ its parity. Any permutation can be expressed in terms
of cyclic exchange processes. Thouless was the first to point out
that cyclic permutations of an even number of spins lead to AF
exchange, whereas when an odd number of spins are permuted the
resulting interaction is FM\cite{Thouless}. Only the values of the
exchange parameters $J_\lambda$ depend on the choice of model; the
form of interaction between spins is quite general.

In La$_2$CuO$_4$, retaining the most important exchange processes
involved in a plaquette, the general effective spin Hamiltonian is
given by:
\begin{eqnarray}\label{eqn2}
{\cal H}_{eff}=J_2^{(1)}\sum_{<ij>}^{(1)} {\cal P}_{ij}^\sigma +
J_2^{(2)}\sum_{<ij>}^{(2)} {\cal P}_{ij}^\sigma +
J_2^{(3)}\sum_{<ij>}^{(3)} {\cal P}_{ij}^\sigma \nonumber\\
-J_3\sum_{ijk}\left[ {\cal P}_{ijk}^\sigma+ \left({\cal
P}_{ijk}^\sigma\right)^{-1}\right]+ J_4\sum_{ijkl}\left[ {\cal
P}_{ijkl}^\sigma+ \left({\cal
P}_{ijkl}^\sigma\right)^{-1}\right]
\end{eqnarray}
the $J_2^{(n)}$ are pair-exchange frequencies between nearest (1),
next-nearest (2) and next-next nearest neighbours (3), $J_3$ and
$J_4$ represent three-and four-particle cyclic exchanges in a
plaquette. In terms of spin operators
\begin{eqnarray}\label{eqn3}
{\cal H}_{eff}= (2J_2^{(1)}-8J_3+2J_4)\sum_{<ij>}^{(1)}{\bf S}_{i}\cdot {\bf S}_{j} \nonumber \\
+ (2J_2^{(2)}-4J_3+J_4)\sum_{<ij>}^{(2)}{\bf S}_{i}\cdot {\bf S}_{j} +
2J_2^{(3)}\sum_{<ij>}^{(3)}{\bf S}_{i}\cdot {\bf S}_{j} \nonumber \\
+ 4J_4\sum_{<ijkl>}\Bigl[ ({\bf S}_i \cdot {\bf S}_j)({\bf S}_k \cdot {\bf S}_l)+
({\bf S}_j \cdot {\bf S}_k)({\bf S}_l\cdot {\bf S}_i) \nonumber \\
- ({\bf S}_{i}\cdot {\bf S}_{k})({\bf S}_{j}\cdot {\bf S}_{l})\Bigr]
\end{eqnarray}

We note that the four-particle cyclic exchange $J_4$ in
Eq.~(\ref{eqn2}) contributes both four-spin and two-spin terms to
Eq.~(\ref{eqn3}). On a square lattice, with the  two-sublattice
antiferromagnetic N\'{e}el phase there is a remarkable (although
fortuitous) exact cancellation of all contributions of the $J_4$
terms in linear spin-wave theory. This means that all quantities (in
particular the magnon dispersion) calculated within this simple
framework are the same as those corresponding to the pure Heisenberg
Hamiltonian
\begin{equation}\label{eqn5}
H_{Heis}= \sum_{n=1}^{n=3}2\tilde J_2^{(n)}\sum_{<ij>}^{(n)}{\bf
S}_i. {\bf S}_j
%
% 2\tilde J_2^{(2)} \sum_{<ij>}^{(2)}{\bf S}_i. {\bf S}_j \nonumber\\
% +2\tilde J_2^{(3)} \sum_{<ij>}^{(2)}{\bf S}_i. {\bf S}_j
\end{equation}
with $\tilde J_2^{(1)}=J_2^{(1)}-4J_3$, $\tilde
J_2^{(2)}=J_2^{(2)}-2J_3$ and $\tilde J_2^{(3)}=J_2^{(3)}$,  and
are completely blind to the four-particle permutation term $J_4$.

For simplicity, we can model the Cu-O planes in La$_2$CuO$_4$
using the half-filled one-band Hubbard model
\begin{equation}\label{eqn4}
H=-t\sum_{ij\sigma}c^+_{i\sigma}c_{j\sigma}+U\sum_i n_\uparrow
n_\downarrow
\end{equation}
where the hopping energy $t$ characterizes the kinetic energy, the
potential energy $U\gg t$ is the penalty for double occupancy, $c$
($c^+$) are the annihilation (creation) operators and $n=c^+ c$ is
a number operator. At fourth order in a $\kappa=t/U$ expansion,
the $J_\lambda$'s appear as $J_2^{(1)}/U=2\kappa^2(1+4\kappa^2)$,
$J_2^{(2)}/U=12\kappa^4$, $J_2^{(3)}/U=2\kappa^4$,
$J_3/U=10\kappa^4$ and $J_4/U=20\kappa^4$\cite{Takahashi,ISSP}.
More intricate expressions are obtained for a more general
three-band Hubbard model\cite{RD}. The effective interaction
between next-nearest neigbour pairs $\tilde J_2^{(2)}$ becomes
negative (i.e. FM) because of the presence of the FM
three-particle term $J_3$. The next-next-nearest neighbour term is
small and can be neglected. The magnitude of the four-particle
cyclic exchange $J_4$ is large.

Since spin waves are insensitive to four-particle cyclic exchange,
the curvature of the magnon dispersion at the zone
boundary\cite{Coldea} is instead entirely due to the ferromagnetic
effective next-nearest neighbour exchange $\tilde J_2^{(2)}$. In
contrast, there is no such cancellation of the four-spin term for
the static susceptibility at high temperatures. We have,
therefore, studied the diffuse scattering in the paramagnetic
phase, and this is a new approach to the investigation of higher
order exchange. The dynamical structure factor for neutron
scattering is given by
\begin{eqnarray}\label{eqn6}
S({\bf Q},\omega)={\omega\over 1-{\rm e}^{-\omega /T}} {S(0)\over
1+(q\xi)^2}\times \nonumber\\
\left[{\Gamma\over (\omega-cq)^2+\Gamma^2} + {\Gamma\over
(\omega+cq)^2+\Gamma^2}\right]
\end{eqnarray}
where $\Gamma$ is the characteristic energy. Integration over energy
transfer yields information on the equal-time spin-spin correlations
since
\begin{equation}\label{eqn7}
\int_{-\infty}^{\infty} S({\bf Q},\omega)d\omega \approx
\sum_i{\rm e}^{\imath {\bf Q \cdot}{\bf R}_i}\langle S_{i}^{z}
S_{0}^{z}\rangle \approx T\chi ({\bf Q}).
\end{equation}
The wave-vector dependent static susceptibility $\chi({\bf Q})$
can be calculated from the exchange energies  using a
high-temperature series expansion.

\begin{figure}[!htb]
  \begin{center}
   \includegraphics[scale=0.4,angle=0]{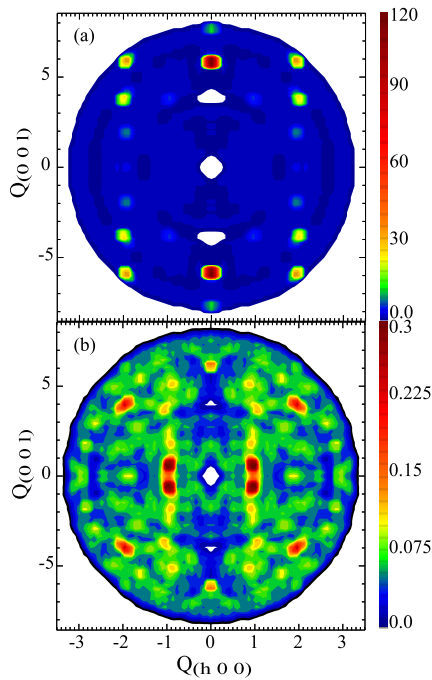}
   \caption{(Color online) Neutron scattering intensity in the $(h,0,l)$ plane of
La$_2$CuO$_4$ at room temperature, i.e. just above the N\'eel
temperature, measured using the multi-detector on D7.
Three-directional polarization analysis allows separation into (a)
coherent structural scattering and (b) purely magnetic scattering,
with removal of the incoherent background. A rod of magnetic
scattering is developing along the $[1,0,l]$ direction showing the
cross-over to 2D correlations.}
   \label{fig:Fig1}
  \end{center}
 \end{figure}

\begin{figure}[!htb]
  \begin{center}
   \includegraphics[scale=0.50,angle=0]{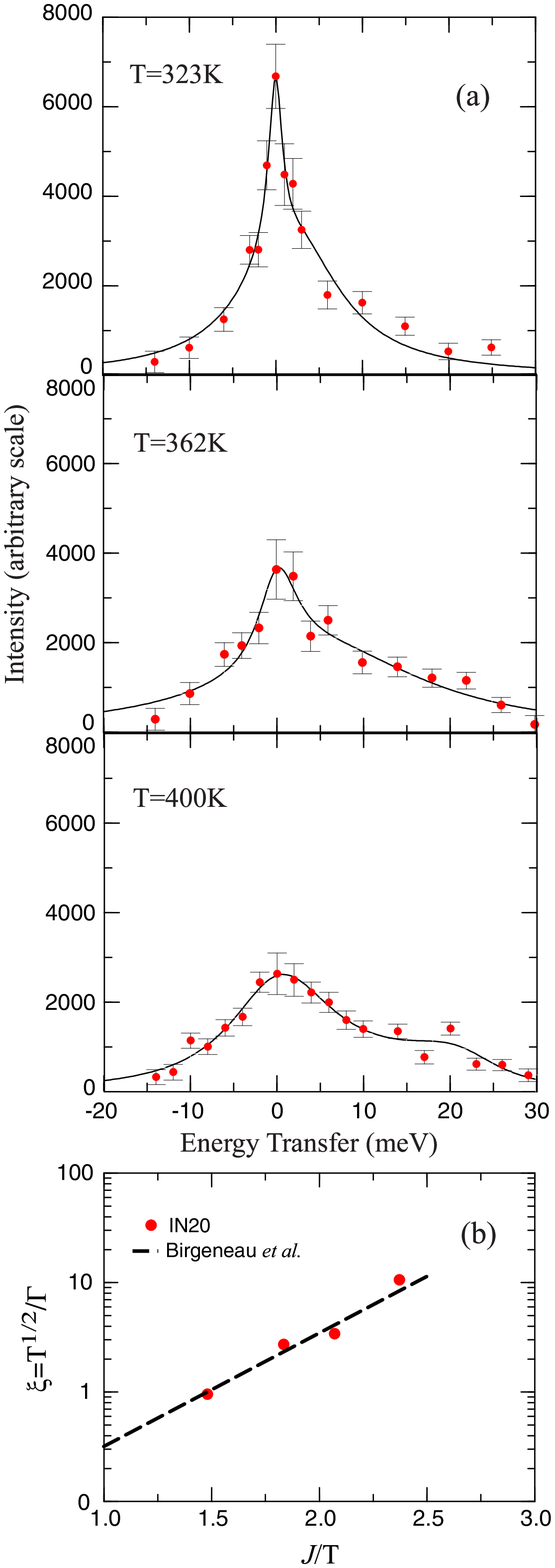}
   \caption{(Color online) (a) Scans of energy transfer at fixed
wave-vector transfer at several temperatures on IN20, and in (b) the
characteristic energies are compared with the correlation lengths
from Ref.\cite{Birgeneau} using the QNL$\sigma$M\cite{Chakravarty}.}
   \label{fig:Fig2}
  \end{center}
 \end{figure}

A $2$g single-crystal of La$_2$CuO$_4$ from the array used to
study the spin waves in the ordered phase\cite{Coldea} was mounted
inside furnaces, and the diffuse magnetic scattering was measured
in the temperature range $300-500$~K using the polarized neutron
spectrometers D7 and IN20 at the Institut Laue-Langevin. $XYZ$
polarization analysis was employed to separate the magnetic signal
from the coherent structural and spin-incoherent
backgrounds\cite{Scharpf}. The scattering intensity measured in
the $(h,0,l)$ plane at room temperature using the multidetector on
D7 is presented in Fig.~\ref{fig:Fig1}(a) for the nuclear
scattering showing the structural Bragg reflections and (b) for
the purely magnetic signal showing the appearance of a rod of
intensity perpendicular to the cuprate planes. The integrated
intensity along the ${\bf Q}^{3D}=(1,0,l)$ rod for La$_2$CuO$_4$
was measured as a function of temperature with the incident wave
vector fixed, $k_i=2.08$\AA$^{-1}$, and the final wave vector
parallel to the normal to the cuprate planes in a similar manner
to Ref.\cite{Birgeneau} so that the cuprate square-lattice wave
vector remained fixed at ${\bf Q}^{2D}=(\frac{1}{2},\frac{1}{2})$
for all energy transfers. For a quantitative temperature
dependence of the intensity integrated over energy transfer it is
essential to determine how the spectral line shape varies with
temperature. Energy scans were performed with ${\bf Q}$ fixed
using the triple-axis spectrometer IN20 and typical spectra are
presented in Fig.~\ref{fig:Fig2}(a).

\begin{figure}[!htb]
  \begin{center}
   \includegraphics[scale=0.48,angle=0]{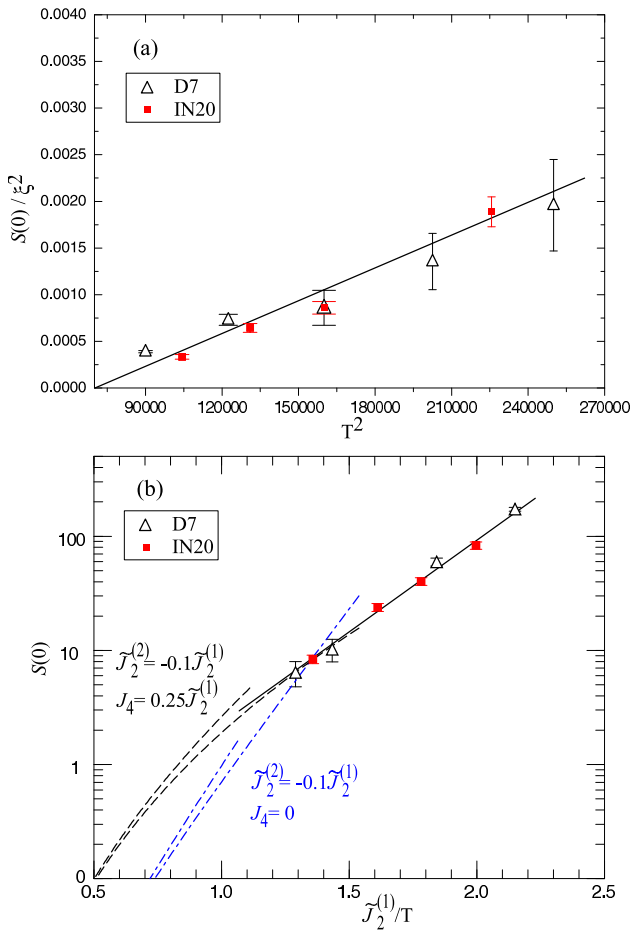}
   \caption{(Color online) The neutron scattering amplitude
$S(0)$ at ${\bf Q}=(\frac{1}{2},\frac{1}{2})$ of the cuprate
square lattice. (a) The ratio of $S(0)$ over the square of the
magnetic correlation length \cite{Birgeneau} varies linearly with
temperature squared, in agreement with the
QNL$\sigma$M\cite{Chakravarty}. (b) The temperature dependence of
$S(0)$ follows a straight (solid) line and the gradient agrees
with the static susceptibility calculated in high-temperature
series expansions with $J_4=0.25\tilde J_{2}^{(1)}$ (dashed line).
The dash-dot line shows the calculation with $J_4=0$. Fourth and
fifth order expansions are shown, the latter extending to lower
temperature.}
   \label{fig:Fig3}
  \end{center}
 \end{figure}

In the QNL$\sigma$M, the correlation length $\xi$ is given
by\cite{Chakravarty}
\begin{equation}
\xi (T)=C_\xi\left[ \frac{\hbar v_s}{\rho_s}\right] \exp{
\left[\frac{2\pi\rho_s}{k_BT}\right] }
\end{equation}
and the energy width $\Gamma$ is related to $\xi$ by
\begin{equation}\label{eqn9}
\Gamma=C_{\Gamma}v_s\left[\frac{T}{2\pi\rho_s}\right]^{\frac{1}{2}}\frac{1}{\xi}
\end{equation}
where $v_s$ is the spin-wave velocity, $\rho_s$ is the spin
stiffness and $C_\xi$ and $C_\Gamma$ are undetermined constants of
order one. Figure~\ref{fig:Fig2}(b) compares the temperature
dependence of $\xi$ deduced from equation~(\ref{eqn9}) using the
values of $\Gamma$ determined in energy scans. There is excellent
agreement between the dynamical predictions of the QNL$\sigma$M
and the correlation lengths measured using unpolarized neutrons by
Birgeneau et al.\cite{Birgeneau}. The intensities measured in
fixed-${\bf Q}$ energy scans on IN20 integrated over energy
transfer were converted to the amplitude S(0) in
equation~(\ref{eqn6}) using the known correlation
lengths\cite{Birgeneau}. The intensities measured on D7 without
energy analysis were corrected using the spectral line shapes
extrapolated from Fig.~\ref{fig:Fig2}(b), the instrumental energy
window, the Cu$^{2+}$ magnetic form factor and the correlation
lengths, and the amplitudes from both experiments are combined in
Fig.~\ref{fig:Fig3}. The leading term in the expression for the
ratio of the amplitude to the correlation length squared in the
QNL$\sigma$M is\cite{Chakravarty}
\begin{equation}
{S(0)\over\xi^2}\approx\left(\frac{k_BT}{2\pi\rho_s}\right)^2.
\end{equation}

Figure~\ref{fig:Fig3}(a) shows that when data collected on D7 are
corrected with a full knowledge of the spectral line shape they
follow the same curve as those collected at fixed ${\bf Q}$ on
IN20. Furthermore, the temperature dependence of intensities
obtained using polarized neutrons is now in agreement with the
predictions of the QNL$\sigma$M. The insensitivity of linear
spin-wave theory to four-particle terms means that the expansion
of four-spin exchange operators in terms of gradients of the
N\'{e}el vector does not add any new terms to the QNL$\sigma$M. It
is, therefore, gratifying that the clean measurements of the
diffuse magnetic signal using polarized neutrons agree now with
the predictions of the QNL$\sigma$M for the renormalized-classical
phase.

In Fig.~\ref{fig:Fig3}(b) we show the comparison of the measured
$S(0)$ with the results of a high-temperature series expansion of
the multiple-spin exchange model Eq.~(\ref{eqn2}).
The high-temperature series expansions were taken to fifth order
and analytically continued using biased Pad\'{e}
approximants\cite{Roger1998}. The values of the pair exchange
energies are those corresponding to the effective pair exchange
$2\tilde J_2^{(1)}=111.8$~meV and $2\tilde J_2^{(2)}=-11.4$~meV
deduced from the magnon spectrum in the ordered
phase\cite{Coldea}. Plotted on a semi log scale to extract the
leading behaviour in $1/T$, the experimental results fall on a
straight line (solid line), and the gradient is in perfect
agreement with the predictions of the series expansion at high
temperatures with $J_4=0.25\tilde J_2^{(1)}$ (dashed line) derived
from the Hubbard model. This agreement is achieved with no free
parameters except an overall scale factor. The dramatic difference
in slope with respect to the dash-dot theoretical line (obtained
with $J_4=0$) demonstrates the extreme sensitivity of the diffuse
magnetic scattering to this term. These data constitute the first
quantitative evidence for four-spin cyclic exchange in
La$_2$CuO$_4$. We note that the ratio $J_4/\tilde J_2^{(1)}\approx
0.25$ is compatible with the predictions of the one-band Hubbard
model, but more accurate neutron data would allow comparison with
a more general three-band model\cite{RD}.

The higher order terms found to be of crucial importance in the
physics of solid $^3$He are also shown to be significant in an
electronic magnetic material. It seems highly likely that ring
exchange will be important in many other electronic magnetic
systems, especially in those with strong hybridisation paths, such
as the Cu$_4$O$_4$ plaquettes. Optical experiments indicate that
higher order exchange is important in other high-temperature
superconductors including YBa$_2$Cu$_3$O$_{6.2}$,
 Bi$_2$Sr$_2$Ca$_{0.5}$Y$_{0.5}$Cu$_2$O$_{8+y}$, Nd$_2$CuO$_4$ and
Pr$_2$CuO$_4$\cite{Sugai}. The magnitude of the four-spin cyclic
exchange is comparable to the pairing energies, and it is possible
that circulating electronic currents have an important role in the
mechanism of superconductivity. Ring exchange is also believed to
be important in related ladder compounds, such as
La$_6$Ca$_8$Cu$_{24}$O$_{41}$ and
Sr$_{14}$Cu$_{24}$O$_{41}$\cite{Brehmer}.

In summary, four-spin cyclic exchange has been resolved in diffuse
scattering experiments in the paramagnetic phase of La$_2$CuO$_4$,
and the 2D critical fluctuations are correctly described by the
QNL$\sigma$M.

We would like to thank S.M. Hayden for the loan of the crystal and
R. Coldea for helpful discussions. Financial support is gratefully
acknowledged from the fifth European Community Framework Programme
through contract HPRN-CT 2000-00166.
%

%\bibliography{ref}
\end{document}